\documentclass[preprint,preprintnumbers,amsmath,amssymb,superscriptaddress]{revtex4-2}

\usepackage{amsthm,amsmath,amsfonts,bm,times,color,bbm}
\usepackage{graphicx}
\usepackage{mhchem}
\usepackage[english]{babel}
\usepackage{mathtools}
\usepackage{float}
\usepackage{subcaption}
\usepackage{hyperref}
\usepackage{multirow}
\usepackage{booktabs}
\usepackage{array}
\usepackage{xcolor}
\usepackage{algorithm}%
\usepackage[font=small,skip=5pt]{caption}
\begin{document}

\title{Long-term prediction of ENSO with physics-guided Deep Echo State Networks}
\author{Zejing Zhang}
\affiliation{School of Physical Science and Technology, Beijing University of Posts and Telecommunications, Beijing, China}

\author{Jun Meng}
\email{jun.meng.phy@gmail.com}
\affiliation{State Key Laboratory of Earth System Numerical Modeling and Application, Institute of Atmospheric Physics, Chinese Academy of Sciences, Beijing, China}

\author{Zhongpu Qiu}
\affiliation{School of Systems Science/Institute of Nonequilibrium Systems, Beijing Normal University, Beijing, China}%

\author{Wansuo Duan}
\affiliation{State Key Laboratory of Earth System Numerical Modeling and Application, Institute of Atmospheric Physics, Chinese Academy of Sciences, Beijing, China}

\author{Jian Gao}
\affiliation{School of Physical Science and Technology, Beijing University of Posts and Telecommunications, Beijing, China}
\affiliation{State Key Laboratory of Information Photonics and Optical Communications, Beijing University of Posts and Telecommunications, Beijing, China}

\author{Zixiang Yan}
\affiliation{School of Physical Science and Technology, Beijing University of Posts and Telecommunications, Beijing, China}

\author{Jinghua Xiao}
\email{jhxiao@bupt.edu.cn}
\affiliation{School of Physical Science and Technology, Beijing University of Posts and Telecommunications, Beijing, China}

\author{Xiaosong Chen}
\affiliation{School of Systems Science/Institute of Nonequilibrium Systems, Beijing Normal University, Beijing, China}

\author{Wenju Cai}
\affiliation{Frontier Science Center for Deep Ocean Multispheres and Earth System (FDOMES) and Physical Oceanography Laboratory, Ocean University of China, Qingdao, China}
\affiliation{Laoshan Laboratory, Qingdao, China}
\affiliation{State Key Laboratory of Marine Environmental Science and College of Ocean and Earth Sciences, Xiamen University, Xiamen, China}
\affiliation{State Key Laboratory of Loess and Quaternary Geology, Institute of Earth Environment, Chinese Academy of Sciences, Xi’an, China}

\author{J\"urgen Kurths}
\affiliation{Potsdam Institute for Climate Impact Research, Potsdam 14412, Germany}
\affiliation{Institute of Physics, Humboldt-University, Berlin, Germany}

\author{Shlomo Havlin}
\affiliation{Department of Physics, Bar-Ilan University, Ramat-Gan, Israel}

\author{Jingfang Fan}
\email{jingfang@bnu.edu.cn}
\affiliation{School of Systems Science/Institute of Nonequilibrium Systems, Beijing Normal University, Beijing, China}
\affiliation{Potsdam Institute for Climate Impact Research, Potsdam 14412, Germany} 

\begin{abstract}
The El Niño–Southern Oscillation (ENSO) is a dominant mode of interannual climate variability, yet the mechanisms limiting its long-lead predictability remain unclear. Here we develop a physics-guided Deep Echo State Network (DESN) that operates on physically interpretable climate modes selected from the extended recharge oscillator (XRO) framework. DESN achieves skillful Niño~3.4 predictions up to 16–20 months ahead with minimal computational cost. Mechanistic experiments show that extended predictability arises from nonlinear coupling between warm water volume and inter-basin climate modes. Error-growth analysis further indicates a finite ENSO predictability horizon of approximately 30 months. These results demonstrate that physics-guided reservoir computing provides an efficient and interpretable framework for diagnosing and predicting ENSO at long lead times.
\end{abstract}

\maketitle

\section{Introduction}
The El Ni\~{n}o–Southern Oscillation (ENSO) is the dominant mode of interannual climate variability, arising from nonlinear ocean–atmosphere interactions in the tropical Pacific and exerting far-reaching impacts on global climate extremes, ecosystems, agriculture, and socio-economic systems~\cite{bib1,bib2,bib3,bib4,bib5,dijkstra_nonlinear_2005}. Classical conceptual models distill the essential physics of ENSO, from the Delayed Oscillator, which emphasizes delayed negative feedbacks associated with equatorial wave reflections~\citep{suarez1988,Battisti1989}, to the Recharge Oscillator (RO), which frames ENSO as a coupled evolution of sea surface temperature and equatorial subsurface heat content governed by Bjerknes feedback, delayed oceanic adjustment, stochastic forcing, and nonlinear atmospheric processes~\citep{bjerknes1969,cane1985,cane1986,jin1997,jinEnsemblemeanDynamicsENSO2007}. State-of-the-art coupled general circulation models reproduce many observed ENSO characteristics and achieve skillful forecasts up to approximately one year~\citep{Barnston2012,Guilyardi2020}.

However, accumulating evidence suggests that ENSO dynamics are embedded within a broader, multiscale inter-basin framework involving the Indian Ocean, the extratropical Pacific, and Atlantic variability~\citep{Izumo2010,Larson2014,Zhang2014,Ham2013,Ding2012,RodriguezFonseca2009,cai2019pantropical}. Data-driven approaches based on network theory and complexity science further demonstrate that dynamical coupling within the tropical Pacific and with remote oceans provides robust precursors of ENSO evolution, enabling forecasts beyond the seasonal timescale~\cite{ludescher2013improved,meng2020complexity}. These insights are formalized in the Extended Recharge Oscillator (XRO), which elevates ENSO from a basin-confined oscillator to a multi-basin coupled dynamical system and extends predictive skill to roughly 18 months~\citep{zhao2024}.

While XRO represents a major advance toward physically interpretable long-lead ENSO prediction, existing conceptual models are formulated in a reduced state space with prescribed feedbacks, emphasizing physical interpretability and diagnostic clarity. In parallel, deep learning (DL) models have demonstrated substantial potential for long-range ENSO prediction by flexibly extracting complex spatiotemporal patterns from high-dimensional climate data. Landmark studies~\cite{ham2019} demonstrated that purely data-driven convolutional networks could rival physical models, and recent transformer-based architectures such as 3D-Geoformer~\cite{zhou2023} have extended Ni\~{n}o3.4 forecast skill beyond 18 months. Despite their predictive success, these architectures are computationally demanding and often difficult to interpret physically, which complicates mechanistic diagnosis, robustness assessment, and regime transfer.

These contrasting strengths raise a central challenge: can machine learning, when guided by physically motivated conceptual frameworks such as XRO, achieve strong long-lead predictive skill while also providing a physically grounded understanding of ENSO dynamics? 
Here, “physically grounded understanding” refers to identifying the interaction pathways, timescales, and nonlinear structures that are dynamically essential for predictability, rather than establishing strict physical causality.

To address this challenge, we develop a physics-guided and computationally efficient forecasting framework based on Deep Echo State Networks (DESNs), a class of reservoir computing models well suited for nonlinear dynamical systems~\cite{jaeger2004,jaeger2007,gallicchio2017,gallicchio2018}. 
Unlike conventional deep learning architectures, DESN combines hierarchical reservoir dynamics with physically interpretable inputs, enabling both long-lead prediction and systematic dynamical diagnosis. 
The model incorporates seasonal priors and ten ENSO-related climate indices selected by the XRO framework~\cite{zhao2024}, allowing nonlinear cross-basin interactions to be represented without prescribing explicit coupling structures.

The proposed DESN framework achieves Ni\~{n}o3.4 forecast anomaly correlation coefficients exceeding 0.5 at lead times of 16–20 months, comparable to state-of-the-art physical and deep learning models~\cite{ham2019,zhou2023,zhao2024}, while requiring only seconds of training on a standard CPU. Beyond predictive skill, DESN supports mechanistic diagnosis through sub-model and controlled experiments, revealing that nonlinear, state-dependent couplings—particularly those involving warm water volume and cross-basin interactions—play a central role in sustaining predictability beyond canonical recharge–discharge dynamics. Building on the error-growth framework underlying nonlinear local Lyapunov exponent (NLLE) analyses~\cite{li2022new,houAsymmetryPredictabilityLimit}, we examine the saturation behavior of forecast errors using an absolute-error metric. The resulting error growth dynamics indicate an intrinsic ENSO predictability horizon of approximately 30 months.

Rather than prioritizing forecast accuracy alone, this work demonstrates that incorporating physically interpretable structure into lightweight recurrent models enables interpretable and theory-consistent ENSO prediction. By guiding a Deep Echo State Network with climate modes motivated by the XRO framework, we show how nonlinear coupling structures and the intrinsic predictability limits imposed by ENSO dynamics jointly shape forecast skill across lead times.

\section{Results}\label{sec2}
To explore long-lead ENSO predictability, we develop a physics-guided DESN framework that operates on a compact set of physically interpretable climate predictors. Guided by the extended recharge oscillator (XRO) framework~\cite{zhao2024}, the model incorporates Niño3.4 (sea surface temperature anomalies averaged over 170°–120°W, 5°S–5°N)~\cite{trenberth1997definition}, warm water volume (WWV), defined as $20^\circ C$ isotherm depth anomalies averaged over 120°E–80°W, 5°S–5°N~\cite{jin1997equatorial}, and a suite of key inter-basin climate modes. These include the North and South Pacific Meridional Modes (NPMM and SPMM)~\cite{chiang2004analogous,zhang2014south}, the Indian Ocean Basin (IOB) mode~\cite{jin2023indian}, the Indian Ocean Dipole (IOD)~\cite{izumo2010influence}, the Southern Indian Ocean Dipole (SIOD)~\cite{jo2022southern}, as well as Tropical North Atlantic (TNA) variability~\cite{ham2013sea}, the Atlantic Niño (ATL3)~\cite{ham2013two}, and the South Atlantic Subtropical Dipole (SASD)~\cite{ham2021inter}, all derived from the ORAS5 ocean reanalysis (see Methods A and Table S1). Together, these ten physically motivated climate indices, along with sinusoidal seasonal cycles, constitute the model inputs. Stochastic variability is incorporated as part of the model’s dynamical evolution and addressed through ensemble forecasting to extract the predictable signal (Fig.~\ref{fig1}b). During inference, the model generates rolling monthly forecasts up to 21 months ahead, initialized at each calendar month with 20-member ensembles. Forecast skill is evaluated using in-sample, out-of-sample, and 24-year block cross-validation schemes (see Methods B–C), providing a robust basis for assessing long-lead ENSO predictability and its dynamical origins.

\subsection{DESN achieves high skill and long-term ENSO forecasts}\label{subsec1}
We trained our DESN models on the ORAS5 dataset over the period 1958–1999 and evaluated them on independent data from 2002–2023, following an out-of-sample forecasting protocol (Supplementary Algorithms, and Table S2). Forecast skill was assessed using the anomaly correlation coefficient (ACC) of the Ni\~{n}o 3.4 index across a range of lead times. As benchmarks, we compared DESN against (1) a minimal Echo State Network (ESN), a standard single-layer reservoir computing model trained only on the Niño~3.4 index,(2) a full ESN trained on ten climate modes (Fig.~\ref{fig1}a) and seasonal cycles, (3) the XRO model~\cite{zhao2024}, (4) operational ensemble forecasts from the International Research Institute for Climate and Society (IRI), and (5) the state-of-the-art 3D-Geoformer deep learning model~\cite{zhou2023}. While all models performed similarly at short lead times, DESN showed clear advantages beyond 10 months, maintaining ACC values above 0.5 up to 16 months (Fig.~\ref{fig:forecast}a). It consistently outperformed the IRI dynamical ensemble beyond 9 months and surpassed the statistical ensemble after just 6 months. 

To evaluate seasonal sensitivity, we computed target-month-dependent forecast skill using rolling initializations. The minimal ESN maintained ACC $>$ 0.5 only at short leads across all months (Fig.~\ref{fig:forecast}c). The full ESN extended this horizon to about 15 months between January and June (Fig.~\ref{fig:forecast}d), but its performance dropped sharply around June and July, reflecting the Spring Predictability Barrier (SPB)~\cite{mcphaden_21st_2012}. In contrast, DESN preserved elevated skill through this season. For example, forecasts targeting June achieved a 10-month skillful lead with DESN compared to only 6 months for the ESN (Fig.~\ref{fig:forecast}e). These results indicate that extending the predictor set from the minimal to the full ESN highlights the importance of cross-basin coupling as a source of ENSO predictability, whereas the further improvement from the full ESN to DESN arises from the deep architecture’s ability to represent higher-order, state-dependent couplings and multiscale interactions that provide additional predictive information.

We further assessed model generalization over extended periods, applying the same hyperparameters (Table. S2) as in the out-of-sample implementation to ensure consistent dynamical behavior. Evaluation was performed over two temporal periods: 1979 to 2023 for the in-sample experiment (Fig.~\ref{fig:forecast}b) and 1958 to 2023 for cross-validation (see Methods C). In the in-sample test, the minimal ESN rapidly lost skill after 3 months, likely reflecting overfitting. The full ESN sustained ACC values $>$ 0.5 up to 18 months, comparable to the XRO and 3D-Geoformer. By contrast, DESN outperformed all baselines, maintaining ACC $>$ 0.5 across lead times up to 20 months, with particularly strong performance during major El Niño episodes (Figs. S1 and S2). Beyond ENSO, the DESN also delivered skillful long-lead predictions for nine other climate modes, achieving lead times of 9–20 months (Fig. S3). Robustness was further demonstrated through 24-year leave-out cross-validation across the full 1958–2023 window, where DESN retained ACC $>$ 0.5 at lead times of 11–18 months (Fig. S4). In addition to forecast accuracy, DESN offers substantial computational efficiency. It trains in seconds on a standard CPU, with runtimes comparable to the XRO while achieving forecast skill that matches or exceeds both dynamical and deep learning approaches (Table.~\ref{tab1}). These results show that DESN generalizes well across decades, avoids overfitting to particular ENSO regimes or climate states, and provides a computationally efficient pathway for advancing long-lead climate prediction.

\begin{table}[t]
\centering
\caption{\textbf{Comparison of training efficiency across ENSO forecasting models.} Estimated training time and computational requirements for the DESN, ESN, XRO, and 3D-Geoformer models.}
\label{tab1}
\setlength{\tabcolsep}{4pt}
\renewcommand{\arraystretch}{1.3}
\begin{tabular}{lllll}
\hline\hline
\textbf{Models} & \textbf{Training data} & \textbf{Training time} & \textbf{Training on} & \textbf{Months} \\
 & \textbf{(size)} & \textbf{(algorithms)} &  & \textbf{lead} \\
\hline
3D-Geoformer & Gridded monthly sea surface & $\sim$12 h & GPU & 13$^a$ \\
\cite{zhou2023} & wind stress and 3D ocean & (Backpropagation) & (Nvidia V100) & 18$^b$ \\
 & temperature anomalies (84 GB) &  &  & \\
XRO & monthly climate indices & 4 s & CPU & 13$^a$ \\
\cite{zhao2024} & (110 kB) & (Linear-regression) & (Intel i9-14900K) & 18$^c$ \\
ESN & monthly climate indices & 26 s & CPU & 13$^a$ \\
 & (110 kB) & (Ridge-regression) & (Intel i9-14900K) & 18$^c$ \\
\textbf{DESN} & \textbf{monthly climate indices} & \textbf{122 s} & \textbf{CPU} & \textbf{16}$^a$ \\
\textbf{(this study)} & \textbf{(110 kB)} & \textbf{(Ridge-regression)} & \textbf{(Intel i9-14900K)} & \textbf{20}$^c$ \\
\hline\hline
\end{tabular}

\vspace{2mm}
{\footnotesize
$^a$ Out-of-sample evaluation (2002–2023); $^b$ Out-of-sample (1983–2021); $^c$ In-sample (1983–2023).
}
\end{table}

\subsection{Mechanisms shaping long-term ENSO predictability}
Here we identify the dynamical mechanisms that sustain ENSO predictability at extended lead times. Using a suite of targeted and computationally efficient experiments enabled by the DESN framework, we show that long-lead ENSO predictability arises from nonlinear, seasonally modulated cross-basin interactions that are integrated into the subsurface ocean memory through WWV. By systematically varying predictor sets, input dimensionality, and model structure, and by directly comparing DESN with the physics-based XRO model, we disentangle the respective roles of seasonal forcing, cross-basin coupling, and nonlinearity. Together, these analyses reveal how physics-guided constraints and nonlinear interaction pathways jointly shape the limits and sources of long-term ENSO predictability.

\subsubsection{Seasonal cycles as periodic bootstrap sequences}
ENSO exhibits strong phase locking to the seasonal cycle, with events typically initiating in boreal spring and peaking in winter~\cite{jin2019,stuecker2013}. To represent this periodicity, we introduced sinusoidal seasonal functions, or Periodic Bootstrap Sequences (PBS), as additional inputs, defined as,
\begin{equation}
\mathbf{PBS}_t=\begin{bmatrix}
 \sin\omega t\\
 \cos\omega t\\
 \sin 2\omega t\\
 \cos 2\omega t
\end{bmatrix},
\label{Eq.PBS}
\end{equation}
where $\omega=\frac{2\pi}{T}=\frac{\pi}{6}$ is the angular frequency of the annual cycle $(T=12)$~\cite{jin2019}, $t \in [0,L-1]$, and $L$ is the number of time steps in the climate indices. PBS terms are applied during both training and prediction phases, significantly altering the reservoir dynamics and enhancing prediction accuracy (Fig. S5). This demonstrates the value of embedding known seasonal structure directly into the model to capture periodic forcing more effectively.

\subsubsection{Long-lead ENSO predictability emerges from WWV–cross-basin coupling}
We interpret long-term ENSO predictability as arising from the nonlinear coupling between WWV and additional inter-basin climate modes, rather than by any single predictor acting in isolation. This interpretation is supported by targeted experiments (see Methods E). To clarify the physical basis of these experiments, we recall that XRO restricts nonlinear terms to ENSO–WWV coupling and a conditional IOD quadratic contribution. In contrast, DESN does not impose explicit constraints on interaction structure, allowing higher-order and cross-basin nonlinear couplings to emerge implicitly from the data.

\paragraph{Mode-decoupling experiments.}
To diagnose the sources and characteristic timescales of predictability, we conduct controlled mode-decoupling experiments in which one climate mode is removed from the full predictor set while all remaining inputs are held fixed. Figs.~\ref{fig:xro_desn_comparison}a and b present the decoupling results for XRO and DESN, respectively. In both models, removing WWV leads to a pronounced collapse of long-lead forecast skill, confirming its indispensable role in sustaining extended ENSO predictability. By contrast, removing other individual modes while retaining WWV generally affects forecast skill but does not eliminate long-term predictability. Notably, the skill reduction associated with WWV removal is more pronounced in DESN than in XRO (Fig. S6), suggesting that DESN exploits WWV through higher-order nonlinear interactions with other climate modes that are not fully represented in the prescribed XRO framework.

\paragraph{Mode-addition experiments.}
To complement the decoupling analysis, we perform controlled mode-addition experiments that explicitly focus on interactions between the core ENSO variables and individual external climate modes. We define a baseline configuration consisting of Niño~3.4 and WWV, which encapsulates the classical recharge--discharge framework. One additional climate mode from the standard XRO set is then added at a time, and both XRO and DESN are trained under identical input conditions. Figs.~\ref{fig:xro_desn_comparison}c and d show the mode-addition results for XRO and DESN, demonstrating that retaining WWV alone is insufficient to sustain long-lead predictability and that extended forecast skill emerges only when WWV is coupled with additional climate modes. Among these, the NPMM yields consistent improvements in both XRO and DESN, highlighting its robust contribution to extended-range ENSO predictability.

\paragraph{Increasing input-dimension experiments.}
Figs.~\ref{fig:xro_desn_comparison}e and f examine systems of progressively increasing dimensionality, defined by the number of physically distinct climate modes coupled to the core ENSO variables (Niño~3.4 and WWV). Predictive skill increases as additional climate indices are combined with WWV and Niño~3.4, confirming that cross-mode interactions substantially enhance ENSO predictability.

Taken together, these experiments provide complementary evidence for a unified mechanism. Retaining ENSO together with external climate modes while excluding WWV fails to recover long-lead forecast skill, indicating that WWV serves as the primary mediator through which ENSO integrates cross-basin influences at extended lead times. Conversely, retaining WWV alone within the mode-addition framework is insufficient to sustain extended predictability. Extended skill emerges only when WWV is nonlinearly coupled with additional inter-basin climate modes, with the NPMM exerting a particularly robust influence. These results demonstrate that long-lead ENSO predictability is governed primarily by the nonlinear coupling between WWV and inter-basin climate modes.

\subsubsection{Nonlinear cross-basin interactions underpin long-lead ENSO predictability}
A first line of evidence for the importance of nonlinearity emerges from the mode-decoupling and mode-addition experiments, as revealed by the systematic comparison between XRO and DESN (Figs.~S6 and S7). Although both models are driven by the same set of climate modes, DESN consistently outperforms XRO at long lead times, while their skills remain comparable at short leads. This divergence emerges despite XRO already incorporating multiple climate modes and instead reflects a structural difference: XRO contains only a limited number of prescribed nonlinear terms, whereas DESN can represent a much broader class of nonlinear interactions through its hierarchical reservoir architecture. The superior long-lead performance of DESN therefore provides indirect but robust evidence that higher-order nonlinear interactions contribute substantially to extended ENSO predictability.

A second, independent line of evidence is provided by the pronounced seasonality of forecast skill. We perform additional uninitialized forecast experiments in which the initial conditions of selected inter-basin modes are replaced by climatology while retaining all learned interactions (see methods E and Fig. S8). These experiments reveal a strong seasonal modulation of inter-basin memory, consistent with previous findings~\cite{zhao2024}. Notably, under specific combinations of initial and target months, removing the initial conditions of Atlantic or Indian Ocean modes can even improve forecast skill, indicating state-dependent and nonlinear interactions that are absent in linear frameworks. Such behavior further supports the view that long-lead predictability arises from nonlinear, seasonally modulated cross-basin influences rather than from linear superposition.

Finally, we obtain strong mechanistic support by introducing an augmented Sparse-Nonlinear XRO (SN-XRO) model (see methods F). Starting from the standard XRO formulation, we systematically introduce quadratic coupling terms representing multiplicative interactions between Niño3.4, WWV, and external climate modes, with seasonally modulated coefficients estimated via sparse regression~\cite{brunton2016,desilva2020}. When trained and evaluated under the same conditions as XRO, SN-XRO behaves similarly to the standard XRO at short lead times but exhibits a pronounced slowdown in skill degradation at longer leads (approximately 8–19 months). This delayed improvement closely mirrors the behavior observed in DESN, providing strong supporting evidence that enhanced nonlinear cross-basin coupling underlies the extension of the forecast horizon (see Figs. S9-S10).

Taken together, these results indicate that WWV influences ENSO predictability through slow, nonlinear, and seasonally modulated cross-basin interactions. This interpretation is consistent with established ENSO theory, in which inter-basin influences are primarily mediated by atmospheric teleconnections and subsequently integrated into subsurface ocean memory~\cite{dijkstra2005nonlinear, mcphaden2006enso,timmermann2018nino}. The convergence of evidence from model-structure comparisons, seasonal sensitivity experiments, and explicit nonlinear augmentation strongly supports the central role of nonlinearity in sustaining long-lead ENSO predictability.

\subsection{Intrinsic limits of ENSO predictability}\label{subsec3}
Although their methodological foundations differ, a range of contemporary ENSO prediction frameworks—including the physics-based XRO conceptual model, the 3D-Geoformer deep learning system, and our ESN/DESN frameworks converge on a forecast horizon of roughly 15–20 months. Given that ENSO cycles span 2–7 years, this convergence raises a fundamental question: does the plateau in forecast skill reflect model limitations, or does it represent an intrinsic barrier imposed by ENSO’s nonlinear dynamics?

To further quantify predictability, we analyze the saturation behavior of forecast errors using an absolute-error metric, inspired by the error-growth framework underlying the NLLE approach. Specifically, we track the temporal evolution of absolute prediction errors following small perturbations to the initial conditions. Error growth in DESN saturates more slowly than in ESN or XRO (Fig.~\ref{fig:lyapunov}a), confirming enhanced dynamical stability. The practical predictability limit was defined as the time when absolute error reached 95\% of its saturation value, estimated consistently across models as the mean absolute error between months 60–100 of the XRO trajectory. Using this criterion, we obtain predictability horizons of approximately 18 months for ESN, 22 months for XRO, and 34 months for DESN. Sensitivity tests further show that reducing the initial perturbation magnitude delays the approach to saturation but yields only limited extensions of the predictability horizon once the initial error norm falls below 0.1 (Fig.~\ref{fig:lyapunov}b), indicating that the estimated limits are robust (Fig.~\ref{fig:lyapunov}b).

These results indicate that the widely cited 15 to 20 month ceiling is not the absolute barrier but rather lies near the intrinsic predictability limit of $\sim$ 30 months.
DESN’s ability to approach this barrier arises not from statistical overfitting but from its physically informed design, stabilized internal dynamics, and improved representation of multiscale evolution. More broadly, the analysis also points to concrete pathways for improving forecast skill: reducing errors in initial conditions, incorporating more precise multivariate couplings across ocean–atmosphere modes, and embedding physically consistent constraints to better capture delayed or cross-basin feedbacks.

\section{Discussion}\label{sec3}
Our results demonstrate that a physics-guided DESN can extend skillful ENSO forecasts well beyond one year, maintaining ACC above 0.5 for 16–20 months. This performance matches or exceeds the skill of operational explainable and state-of-the-art deep learning models, while requiring only seconds of training on a standard CPU. Such efficiency enables systematic exploration of model configurations and physical mechanisms, bridging the gap between predictive skill and mechanistic understanding.

A central contribution of this study is clarifying the mechanisms that sustain long-lead predictability. Submodel experiments confirm that WWV is the dominant precursor of ENSO, consistent with the recharge–discharge paradigm, but also show that its predictive power is amplified by couplings with remote modes such as the NPMM. These delayed and cross-basin feedbacks extend predictability, and DESN’s hierarchical structure is particularly effective in capturing such multiscale interactions. 

From a broader climate perspective, our results highlight physics-guided learning as a surrogate dynamical model, capable of delivering both forecast skill and mechanistic clarity. By showing how WWV-mediated interactions and multiscale feedbacks shape the intrinsic predictability limit of ENSO, this study advances understanding of ENSO as a nonlinear dynamical system. Looking ahead, integrating improved initialization, richer multivariate observations, and coupled Earth system models may further enhance long-lead forecasts. More broadly, this framework illustrates how machine learning can be embedded within dynamical theory to advance prediction and understanding of other critical components of the climate system.

\bibliography{sn-bibliography}

@article{cai2019pantropical,
  title={Pantropical climate interactions},
  author={Cai, Wenju and Wu, Lixin and Lengaigne, Matthieu and Li, Tim and McGregor, Shayne and Kug, Jong-Seong and Yu, Jin-Yi and Stuecker, Malte F and Santoso, Agus and Li, Xichen and others},
  journal={Science},
  volume={363},
  number={6430},
  pages={eaav4236},
  year={2019},
  publisher={American Association for the Advancement of Science}
}

@article{jin2023indian,
  title={The indian ocean weakens the ENSO spring predictability barrier: role of the indian ocean basin and dipole modes},
  author={Jin, Yishuai and Meng, Xing and Zhang, Li and Zhao, Yingying and Cai, Wenju and Wu, Lixin},
  journal={Journal of Climate},
  volume={36},
  number={24},
  pages={8331--8345},
  year={2023},
  publisher={American Meteorological Society}
}

@article{izumo2010influence,
  title={Influence of the state of the Indian Ocean Dipole on the following year’s El Ni{\~n}o},
  author={Izumo, Takeshi and Vialard, J{\'e}r{\^o}me and Lengaigne, Matthieu and de Boyer Montegut, Cl{\'e}ment and Behera, Swadhin K and Luo, Jing-Jia and Cravatte, Sophie and Masson, S{\'e}bastien and Yamagata, Toshio},
  journal={Nature Geoscience},
  volume={3},
  number={3},
  pages={168--172},
  year={2010},
  publisher={Nature Publishing Group UK London}
}

@article{jo2022southern,
  title={Southern Indian Ocean Dipole as a trigger for central Pacific El Ni{\~n}o since the 2000s},
  author={Jo, Hyun-Su and Ham, Yoo-Geun and Kug, Jong-Seong and Li, Tim and Kim, Jeong-Hwan and Kim, Ji-Gwang and Kim, Hyerim},
  journal={Nature Communications},
  volume={13},
  number={1},
  pages={6965},
  year={2022},
  publisher={Nature Publishing Group UK London}
}

@article{ham2013sea,
  title={Sea surface temperature in the north tropical Atlantic as a trigger for El Ni{\~n}o/Southern Oscillation events},
  author={Ham, Yoo-Geun and Kug, Jong-Seong and Park, Jong-Yeon and Jin, Fei-Fei},
  journal={Nature Geoscience},
  volume={6},
  number={2},
  pages={112--116},
  year={2013},
  publisher={Nature Publishing Group UK London}
}

@article{ham2013two,
  title={Two distinct roles of Atlantic SSTs in ENSO variability: North tropical Atlantic SST and Atlantic Ni{\~n}o},
  author={Ham, Yoo-Geun and Kug, Jong-Seong and Park, Jong-Yeon},
  journal={Geophysical Research Letters},
  volume={40},
  number={15},
  pages={4012--4017},
  year={2013},
  publisher={Wiley Online Library}
}

@article{ham2021inter,
  title={Inter-basin interaction between variability in the South Atlantic Ocean and the El Ni{\~n}o/Southern oscillation},
  author={Ham, Yoo-Geun and Lee, Hyun-Jeong and Jo, Hyun-Su and Lee, Se-Gun and Cai, Wenju and Rodrigues, Regina R},
  journal={Geophysical Research Letters},
  volume={48},
  number={15},
  pages={e2021GL093338},
  year={2021},
  publisher={Wiley Online Library}
}

@article{trenberth1997definition,
  title={The definition of el nino},
  author={Trenberth, Kevin E},
  journal={Bulletin of the American Meteorological Society},
  volume={78},
  number={12},
  pages={2771--2778},
  year={1997},
  publisher={American Meteorological Society}
}

@article{jin1997equatorial,
  title={An equatorial ocean recharge paradigm for ENSO. Part I: Conceptual model},
  author={Jin, Fei-Fei},
  journal={Journal of the atmospheric sciences},
  volume={54},
  number={7},
  pages={811--829},
  year={1997}
}

@article{zhang2014south,
  title={The South Pacific meridional mode: A mechanism for ENSO-like variability},
  author={Zhang, Honghai and Clement, Amy and Di Nezio, Pedro},
  journal={Journal of Climate},
  volume={27},
  number={2},
  pages={769--783},
  year={2014}
}

@article{chiang2004analogous,
  title={Analogous Pacific and Atlantic meridional modes of tropical atmosphere--ocean variability},
  author={Chiang, John CH and Vimont, Daniel J},
  journal={Journal of Climate},
  volume={17},
  number={21},
  pages={4143--4158},
  year={2004}
}

@article{meng2020complexity,
  title={Complexity-based approach for {El Ni{\~n}o} magnitude forecasting before the spring predictability barrier},
  author={Meng, Jun and Fan, Jingfang and Ludescher, Josef and Agarwal, Ankit and Chen, Xiaosong and Bunde, Armin and Kurths, J{\"u}rgen and Schellnhuber, Hans Joachim},
  journal={Proceedings of the National Academy of Sciences},
  volume={117},
  number={1},
  pages={177--183},
  year={2020},
  publisher={National Academy of Sciences}
}

@article{jinEnsemblemeanDynamicsENSO2007,
  title = {Ensemble-Mean Dynamics of the {{ENSO}} Recharge Oscillator under State-Dependent Stochastic Forcing},
  author = {Jin, Fei-Fei and Lin, L. and Timmermann, A. and Zhao, J.},
  year = 2007,
  journal = {Geophysical Research Letters},
  volume = {34},
  number = {3},
  issn = {1944-8007},
  doi = {10.1029/2006GL027372},
  urldate = {2025-12-27},
}

@article{timmermann2018nino,
  title={El Ni{\~n}o--southern oscillation complexity},
  author={Timmermann, Axel and An, Soon-Il and Kug, Jong-Seong and Jin, Fei-Fei and Cai, Wenju and Capotondi, Antonietta and Cobb, Kim M and Lengaigne, Matthieu and McPhaden, Michael J and Stuecker, Malte F and others},
  journal={Nature},
  volume={559},
  number={7715},
  pages={535--545},
  year={2018},
  publisher={Nature Publishing Group UK London}
}

@book{dijkstra2005nonlinear,
  title={Nonlinear physical oceanography: a dynamical systems approach to the large scale ocean circulation and El Nino},
  author={Dijkstra, Henk A},
  volume={532},
  year={2005},
  publisher={Springer}
}

@article{mcphaden2006enso,
  title={ENSO as an integrating concept in earth science},
  author={McPhaden, Michael J and Zebiak, Stephen E and Glantz, Michael H},
  journal={science},
  volume={314},
  number={5806},
  pages={1740--1745},
  year={2006},
  publisher={American Association for the Advancement of Science}
}

@article{ludescher2013improved,
  title={Improved {El Ni{\~n}o} forecasting by cooperativity detection},
  author={Ludescher, Josef and Gozolchiani, Avi and Bogachev, Mikhail I and Bunde, Armin and Havlin, Shlomo and Schellnhuber, Hans Joachim},
  journal={Proceedings of the National Academy of Sciences},
  volume={110},
  number={29},
  pages={11742--11745},
  year={2013},
  publisher={National Academy of Sciences}
}

@article{li2022new,
  title={A New Technique to Quantify the Local Predictability of Extreme Events: The Backward Nonlinear Local Lyapunov Exponent Method},
  author={Li, Xuan and Ding, Ruiqiang and Li, Jianping},
  journal={Frontiers in Environmental Science},
  volume={10},
  pages={825233},
  year={2022},
  publisher={Frontiers Media SA}
}

@article{bib1,
  author = {Dake Chen and Stephen E. Zebiak and Antonio J. Busalacchi and Mark A. Cane},
  title = {An Improved Procedure for {El Niño} Forecasting: Implications for Predictability},
  journal = {Science},
  volume = {269},
  number = {5231},
  pages = {1699-1702},
  year = {1995},
  doi = {10.1126/science.269.5231.1699}
}

@article{bib2,
  author = {Michael J. McPhaden and Stephen E. Zebiak and Michael H. Glantz},
  title = {{ENSO} as an Integrating Concept in Earth Science},
  journal = {Science},
  volume = {314},
  number = {5806},
  pages = {1740-1745},
  year = {2006}
}

@article{bib3,
  author = {Timmermann, A. and others},
  title = {{El Niño--Southern Oscillation} Complexity},
  journal = {Nature},
  volume = {559},
  number = {7715},
  pages = {535--545},
  year = {2018},
  month = jul,
  doi = {10.1038/s41586-018-0394-2}
}

@article{bib4,
  author = {Zhang, Rong-Hua and Rothstein, Lewis M. and Busalacchi, Antonio J.},
  title = {Origin of Upper-Ocean Warming and {El Niño} Change on Decadal Scales in the Tropical Pacific Ocean},
  journal = {Nature},
  volume = {391},
  number = {6670},
  pages = {879--883},
  year = {1998},
  doi = {10.1038/36081}
}

@article{bib5,
  author = {Cai, Wenju and others},
  title = {Changing {El Niño--Southern Oscillation} in a Warming Climate},
  journal = {Nature Reviews Earth \& Environment},
  year = {2021},
  doi = {10.1038/s43017-021-00199-z}
}

@article{cane1986,
  author = {Cane, Mark A. and Zebiak, Stephen E. and Dolan, Sean C.},
  title = {Experimental Forecasts of {El Niño}},
  journal = {Nature},
  volume = {321},
  number = {6073},
  pages = {827--832},
  year = {1986},
  month = jun,
  doi = {10.1038/321827a0}
}

@article{cane1985,
  author = {Cane, Mark A. and Zebiak, Stephen E.},
  title = {A Theory for {El Niño} and the Southern Oscillation},
  journal = {Science},
  volume = {228},
  number = {4703},
  pages = {1085--1087},
  year = {1985},
  month = may,
  doi = {10.1126/science.228.4703.1085}
}

@article{jin1997,
  author = {Jin, Fei-Fei},
  title = {An Equatorial Ocean Recharge Paradigm for {ENSO}. Part {I}: Conceptual Model},
  journal = {Journal of the Atmospheric Sciences},
  volume = {54},
  number = {7},
  pages = {811--829},
  year = {1997},
  month = apr,
  doi = {10.1175/1520-0469(1997)054<0811:AEORPF>2.0.CO;2}
}

@article{suarez1988,
  author = {Suarez, Max J. and Schopf, Paul S.},
  title = {A Delayed Action Oscillator for {ENSO}},
  journal = {Journal of the Atmospheric Sciences},
  volume = {45},
  number = {21},
  pages = {3283--3287},
  year = {1988},
  month = nov,
  doi = {10.1175/1520-0469(1988)045<3283:ADAOFE>2.0.CO;2}
}

@article{ham2019,
  title = {Deep Learning for Multi-Year {ENSO} Forecasts},
  author = {Ham, Yoo-Geun and Kim, Jeong-Hwan and Luo, Jing-Jia},
  journal = {Nature},
  volume = {573},
  number = {7775},
  pages = {568--572},
  year = {2019},
  doi = {10.1038/s41586-019-1559-7}
}

@article{zhou2023,
  title = {A Self-Attention--Based Neural Network for Three-Dimensional Multivariate Modeling and Its Skillful {ENSO} Predictions},
  author = {Zhou, Lu and Zhang, Rong-Hua},
  journal = {Science Advance},
  volume = {9},
  number = {10},
  pages = {eadf2827},
  year = {2023},
  doi = {10.1126/sciadv.adf2827}
}

@article{zhao2024,
  title = {Explainable {El Niño} Predictability from Climate Mode Interactions},
  author = {Zhao, Sen and Jin, Fei-Fei and Stuecker, Malte F. and Thompson, Philip R. and Kug, Jong-Seong and McPhaden, Michael J. and Cane, Mark A. and Wittenberg, Andrew T. and Cai, Wenju},
  journal = {Nature},
  volume = {630},
  number = {8018},
  pages = {891--898},
  year = {2024},
  doi = {10.1038/s41586-024-07534-6}
}

@article{gallicchio2017,
  title = {Deep Reservoir Computing: A Critical Experimental Analysis},
  author = {Gallicchio, Claudio and Micheli, Alessio and Pedrelli, Luca},
  journal = {Neurocomputing},
  volume = {268},
  pages = {87--99},
  year = {2017},
  doi = {10.1016/j.neucom.2016.12.089}
}

@article{gallicchio2018,
  title = {Design of Deep Echo State Networks},
  author = {Gallicchio, Claudio and Micheli, Alessio and Pedrelli, Luca},
  journal = {Neural Networks},
  volume = {108},
  pages = {33--47},
  year = {2018},
  doi = {10.1016/j.neunet.2018.08.002}
}

@article{jaeger2007,
  title = {Echo State Network},
  author = {Jaeger, Herbert},
  journal = {Scholarpedia},
  volume = {2},
  number = {9},
  pages = {2330},
  year = {2007},
  doi = {10.4249/scholarpedia.2330}
}

@article{jaeger2004,
  title = {Harnessing Nonlinearity: Predicting Chaotic Systems and Saving Energy in Wireless Communication},
  author = {Jaeger, Herbert and Haas, Harald},
  journal = {Science},
  volume = {304},
  number = {5667},
  pages = {78--80},
  year = {2004},
  doi = {10.1126/science.1091277}
}

@article{jaegerOptimization2007,
  title = {Optimization and Applications of Echo State Networks with Leaky-Integrator Neurons},
  author = {Jaeger, Herbert and Luko{\v s}evi{\v c}ius, Mantas and Popovici, Dan and Siewert, Udo},
  journal = {Neural Networks},
  volume = {20},
  number = {3},
  pages = {335--352},
  year = {2007},
  doi = {10.1016/j.neunet.2007.04.016}
}

@article{jin2019,
  title = {Seasonal Cycle of Background in the Tropical Pacific as a Cause of ENSO Spring Persistence Barrier},
  author = {Jin, Yishuai and Liu, Zhengyu and Lu, Zhengyao and He, Chengfei},
  journal = {Geophysical Research Letters},
  volume = {46},
  number = {22},
  pages = {13371--13378},
  year = {2019},
  doi = {10.1029/2019GL085205}
}

@article{stuecker2013,
  title = {A Combination Mode of the Annual Cycle and the El Ni{\~n}o/Southern Oscillation},
  author = {Stuecker, Malte F. and Timmermann, Axel and Jin, Fei-Fei and McGregor, Shayne and Ren, Hong-Li},
  journal = {Nature Geoscience},
  volume = {6},
  number = {7},
  pages = {540--544},
  year = {2013},
  doi = {10.1038/ngeo1826}
}

@book{dijkstra_nonlinear_2005,
	edition = {2},
	series = {Atmospheric and {Oceanographic} {Sciences} {Library}},
	title = {Nonlinear {Physical} {Oceanography}: {A} {Dynamical} {Systems} {Approach} to the {Large} {Scale} {Ocean} {Circulation} and {El} {Niño}},
	isbn = {978-1-4020-2262-3},
	issn = {1383-8601}, 
	language = {en},
	publisher = {Springer Netherlands},
	author = {Dijkstra, Henk A.},
	year = {2005},
	pages = {xvi+532}
}

@article{mcphaden_21st_2012,
	title = {A 21st Century Shift in the Relationship Between {ENSO} {SST} and Warm Water Volume Anomalies},
	volume = {39},
	number = {9},
	pages = {L09706},
	issn = {1944-8007},
	journal = {Geophysical Research Letters},
	author = {McPhaden, Michael J.},
	year = {2012},
	doi = {10.1029/2012GL051826}
}

@article{houAsymmetryPredictabilityLimit,
	title = {Asymmetry of the Predictability Limit of the Warm {ENSO} Phase},
	volume = {45},
	number = {23},
	pages = {12947--12955},
	issn = {1944-8007},
	journal = {Geophysical Research Letters},
	author = {Hou, Zhaolu and Li, Jianping and Ding, Ruiqiang and Karamperidou, Christina and Duan, Wansuo and Liu, Ting and Feng, Jie},
	year = {2018},
	doi = {10.1029/2018GL077880}
}

@article{Barnston2012,
  author = {Barnston, Anthony G. and Tippett, Michael K. and L'Heureux, Michelle L. and Li, Shuhua and DeWitt, David G.},
  title = {Skill of Real-Time Seasonal {ENSO} Model Predictions During 2002--11: Is Our Capability Increasing?},
  journal = {Bulletin of the American Meteorological Society},
  volume = {93},
  number = {5},
  pages = {631--651},
  year = {2012},
  doi = {10.1175/BAMS-D-11-00111.1}
}

@article{Battisti1989,
  author = {Battisti, David S. and Hirst, Anthony C.},
  title = {Interannual Variability in a Tropical Atmosphere-Ocean Model: Influence of the Basic State, Ocean Geometry and Nonlinearity},
  journal = {Journal of the Atmospheric Sciences},
  volume = {46},
  number = {12},
  pages = {1687--1712},
  year = {1989},
  doi = {10.1175/1520-0469(1989)046<1687:IVIATA>2.0.CO;2}
}

@article{Bjerknes1969,
  author = {Bjerknes, Jacob},
  title = {Atmospheric Teleconnections from the Equatorial {Pacific}},
  journal = {Monthly Weather Review},
  volume = {97},
  number = {3},
  pages = {163--172},
  year = {1969},
  doi = {10.1175/1520-0493(1969)097<0163:ATFTEP>2.3.CO;2}
}

@article{Ding2012,
  author = {Ding, Hui and Keenlyside, Noel S. and Latif, Mojib},
  title = {Impact of the Equatorial {Atlantic} on the {El Ni\~no Southern} Oscillation},
  journal = {Climate Dynamics},
  volume = {38},
  number = {9-10},
  pages = {1965--1972},
  year = {2012},
  doi = {10.1007/s00382-011-1097-y}
}

@article{Ham2013,
  author = {Ham, Yoo-Geun and Kug, Jong-Seong and Park, June-Yi and Jin, Fei-Fei},
  title = {Sea Surface Temperature in the North Tropical {Atlantic} as a Trigger for {El Ni\~no/Southern Oscillation} Events},
  journal = {Nature Geoscience},
  volume = {6},
  number = {2},
  pages = {112--116},
  year = {2013},
  doi = {10.1038/ngeo1686}
}

@article{Izumo2010,
  author = {Izumo, Takeshi and Vialard, J{\'e}r{\^o}me and Lengaigne, Matthieu and de Boyer Mont{\'e}gut, Cl{\'e}ment and Behera, Swadhin K. and Luo, Jing-Jia and Cravatte, Sophie and Masson, S{\'e}bastien and Yamagata, Toshio},
  title = {Influence of the State of the {Indian Ocean Dipole} on the Following Year's {El Ni\~no}},
  journal = {Nature Geoscience},
  volume = {3},
  number = {3},
  pages = {168--172},
  year = {2010},
  doi = {10.1038/ngeo760}
}

@article{Larson2014,
  author = {Larson, Sarah M. and Kirtman, Ben P.},
  title = {The {Pacific} Meridional Mode as an {ENSO} Precursor and Predictor in the {North American Multimodel Ensemble}},
  journal = {Journal of Climate},
  volume = {27},
  number = {18},
  pages = {7018--7032},
  year = {2014},
  doi = {10.1175/JCLI-D-14-00055.1}
}

@article{RodriguezFonseca2009,
  author = {Rodr{\'i}guez-Fonseca, Bel{\'e}n and Polo, Irene and Garc{\'i}a-Serrano, Javier and Losada, Teresa and Mohino, Elsa and Mechoso, Carlos R. and Kucharski, Fred},
  title = {Are {Atlantic Ni\~nos} Enhancing {Pacific ENSO} Events in Recent Decades?},
  journal = {Geophysical Research Letters},
  volume = {36},
  number = {20},
  pages = {L20705},
  year = {2009},
  doi = {10.1029/2009GL040048}
}

@article{Zhang2014,
  author = {Zhang, Hui and Clement, Amy and Di Nezio, Pedro},
  title = {The {South Pacific Meridional Mode}: A Mechanism for {ENSO}-like Variability},
  journal = {Journal of Climate},
  volume = {27},
  number = {2},
  pages = {769--783},
  year = {2014},
  doi = {10.1175/JCLI-D-13-00082.1}
}

@incollection{Guilyardi2020,
  author = {Guilyardi, Eric and Capotondi, Antonietta and Lengaigne, Matthieu and Thual, Sulian and Wittenberg, Andrew T.},
  title = {{ENSO} Modeling: History, Progress, and Challenges},
  booktitle = {El Ni\~no Southern Oscillation in a Changing Climate},
  editor = {McPhaden, Michael J. and Santoso, Agus and Cai, Wenju},
  pages = {199--226},
  year = {2020},
  publisher = {American Geophysical Union},
  address = {Washington, DC},
  doi = {10.1002/9781119548164.ch9}
}

@article{brunton2016,
  title={Discovering governing equations from data by sparse identification of nonlinear dynamical systems},
  author={Brunton, Steven L and Proctor, Joshua L and Kutz, J Nathan},
  journal={Proceedings of the National Academy of Sciences},
  volume={113},
  number={15},
  pages={3932--3937},
  year={2016},
  publisher={National Acad Sciences},
  doi={10.1073/pnas.1517384113}
}

@article{desilva2020,
  title={PySINDy: A Python package for the sparse identification of nonlinear dynamical systems from data},
  author={de Silva, Brian M and Champion, Kathleen and Quade, Markus and Loiseau, Jean-Christophe and Kutz, J Nathan and Brunton, Steven L},
  journal={Journal of Open Source Software},
  volume={5},
  number={49},
  pages={2104},
  year={2020},
  doi={10.21105/joss.02104}
}

\begin{figure*}[h!] 
    \centering
    \includegraphics[width=1.0\textwidth]{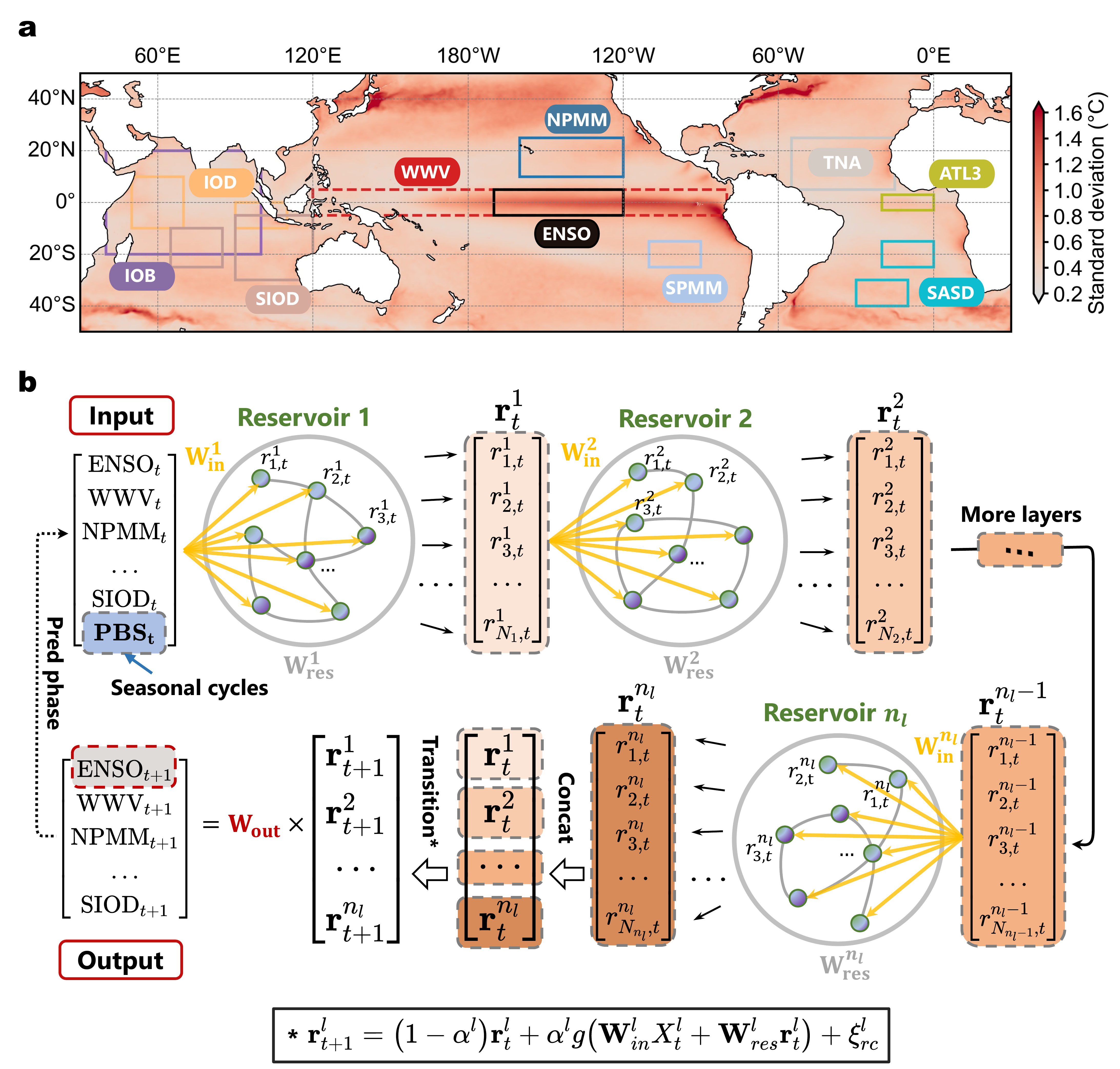}
    \caption{\textbf{Regions of interest and architecture of the DESN model for ENSO prediction.}  \textbf{a}, Standard deviation of sea surface temperature anomalies (SSTA) from the ORAS5 reanalysis (1979–2023). Colored boxes denote regions used to compute area-averaged climate indices for ENSO and associated modes, following the XRO framework~\cite{zhao2024}. 
    \textbf{b}, Schematic of the Deep Echo State Network (DESN) architecture. The input layer comprises multiple climate indices and seasonal cycles as periodic bootstrap sequences ($\mathbf{PBS_t}$). Inputs are passed through a hierarchy of $n_l$ recurrent reservoirs, each generating a neuron state vector $\mathbf{r}_t^l$. Additive white noise $\xi_{\text{rc}}^l$ is injected into each reservoir to emulate stochastic variability. Internal weights consist of input matrices $\mathbf{W}_{\text{in}}^l$ and reservoir matrices $\mathbf{W}_{\text{res}}^l$. Reservoir outputs are concatenated and mapped to next-month targets via a linear readout matrix $\mathbf{W}_{\text{out}}$. Multi-step forecasts are obtained recursively using rolling inputs. When $n_l = 1$, the model reduces to a standard ESN. The reservoir transition function at the bottom illustrates neuron dynamics with activation function $g(\cdot)$ and leakage rate $\alpha$. See Methods for definitions and implementation details.}
    \label{fig1}
\end{figure*}

\begin{figure*} [htbp]
    \centering
    \includegraphics[width=0.9\textwidth]{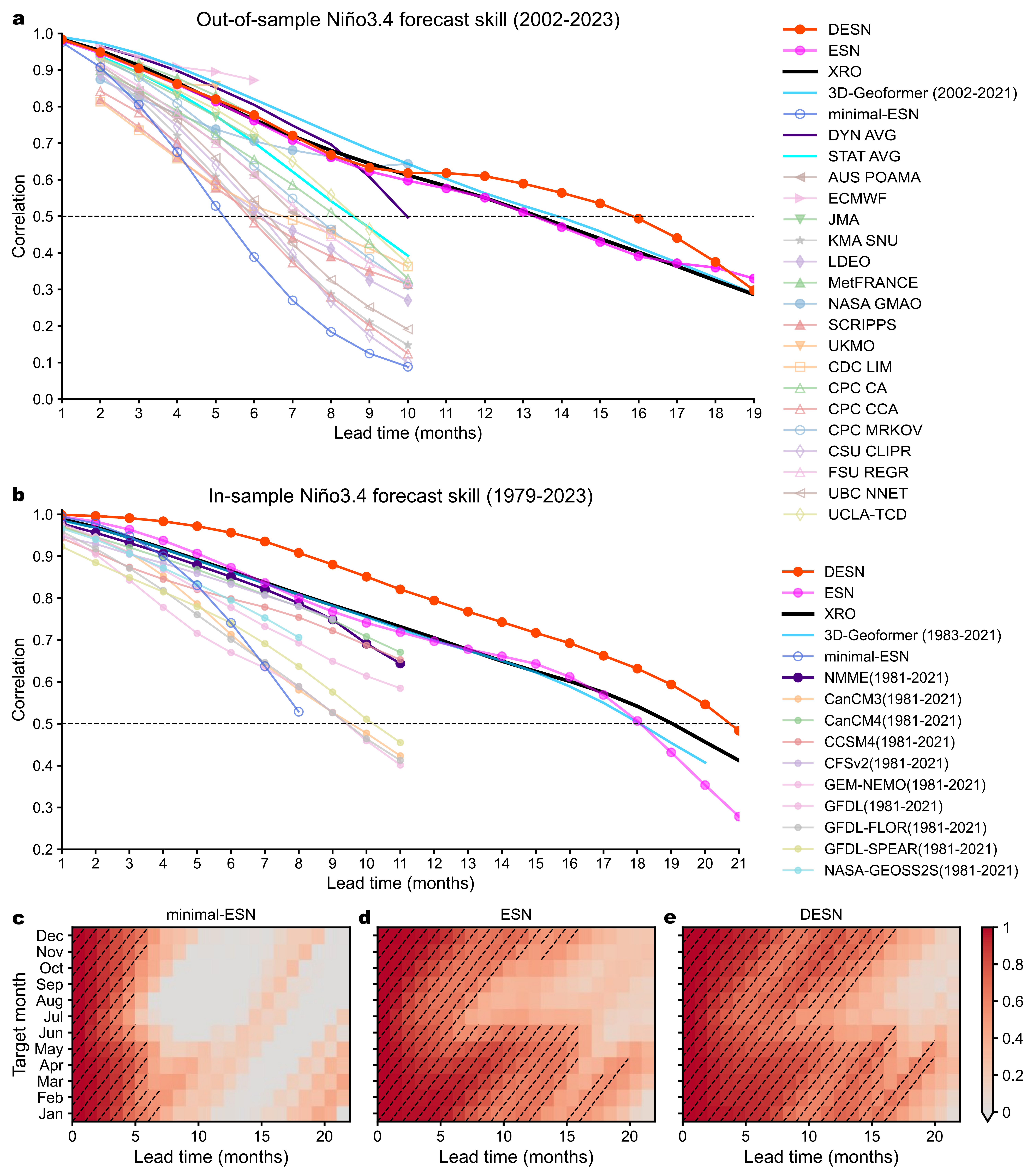}
    \caption{\textbf{Forecast performance of ESN and DESN models for ENSO prediction.} 
    \textbf{a}, Forecast correlation skill (ACC) of the 3-month running mean Ni\~{n}o3.4 index as a function of lead time, evaluated out-of-sample for 2002–2023. Shown are DESN (red), ESN (magenta), minimal-ESN (dark blue), XRO (black; trained on 1958–1999), 3D-Geoformer (light blue), and the IRI operational ensemble forecasts—ensemble means of dynamical models (dark purple) and statistical models (dark cyan). 
   \textbf{b}, Same as panel~\textbf{a}, but evaluated in-sample over 1979-2023, including the ensemble mean of dynamical models from the North American Multi-Model Ensemble (NMME). Individual NMME model forecasts (1981–2021) are shown in various colors. Forecast periods for the 3D-Geoformer and NMME correspond to their respective training spans. 
   \textbf{c-e}, Target-month-dependent ACC of Ni\~{n}o3.4 forecasts for the minimal-ESN (\textbf{c}), ESN (\textbf{d}), and DESN (\textbf{e}). Colors denote correlation skill as a function of target month (vertical axis) and lead time (horizontal axis). Black contours delineate regions where ACC exceeds 0.5, highlighting combinations of target months and lead times with high predictive skill.}
    \label{fig:forecast}
\end{figure*}

\begin{figure*}[h!]
\centering
\includegraphics[width=1.0\textwidth]{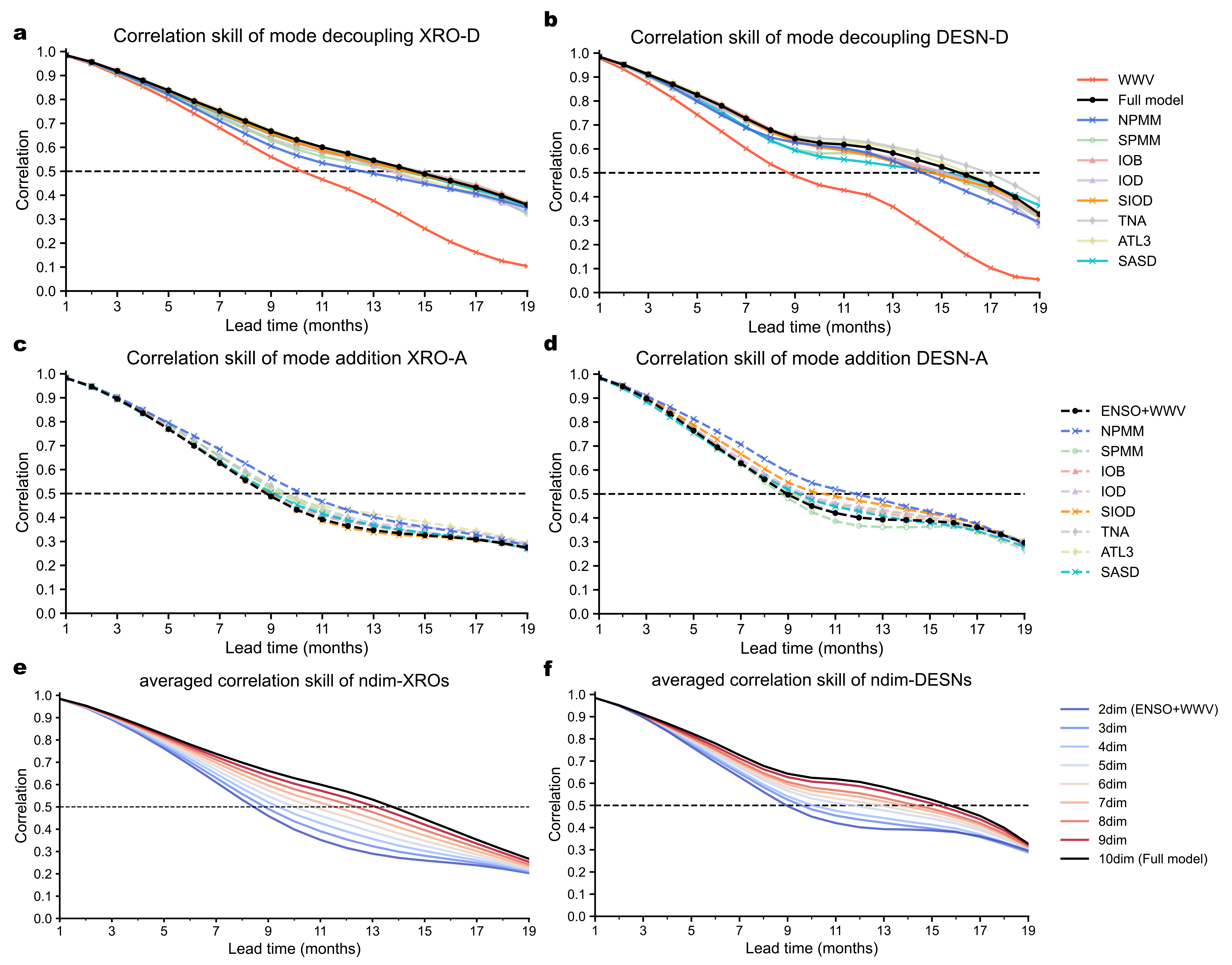}
 \caption{\textbf{Role of WWV and cross-basin modes in long-lead ENSO predictability.}
    \textbf{a,b}, Mode-decoupling experiments for XRO (\textbf{a}) and DESN (\textbf{b}). Each curve shows the anomaly correlation coefficient (ACC) of Niño~3.4 forecasts as a function of lead time when one climate mode is removed from the full predictor set while all remaining inputs are retained. The black curve denotes the full model, and colored curves correspond to decoupling individual modes (WWV, NPMM, SPMM, IOB, IOD, SIOD, TNA, ATL3, and SASD). Removing WWV leads to the strongest degradation of long-lead skill in both models, while removing other modes generally reduces skill without eliminating extended predictability. The horizontal dashed line marks ACC = 0.5.
    \textbf{c,d}, Mode-addition experiments for XRO (\textbf{c}) and DESN (\textbf{d}). Forecast skill is shown for a baseline configuration consisting of Niño~3.4 and WWV (black dashed curve), with one additional climate mode added at a time (colored curves). Extended forecast skill emerges only when WWV is coupled with additional inter-basin modes, with the North Pacific Meridional Mode (NPMM) yielding particularly robust improvements. The dashed horizontal line again indicates ACC = 0.5. 
    \textbf{e,f}, Increasing input-dimensionality experiments for WWV-selected sub-models in XRO (\textbf{e}) and DESN (\textbf{f}). Curves show the average forecast skill across ensembles of sub-models with progressively increasing numbers of climate modes coupled to Niño~3.4 and WWV (from 2 to 9 dimensions). Forecast skill increases with input dimensionality, indicating the cumulative contribution of cross-mode interactions.}
    \label{fig:xro_desn_comparison}
\end{figure*}

\begin{figure*}[htbp] 
    \centering
    \includegraphics[width=1\textwidth]{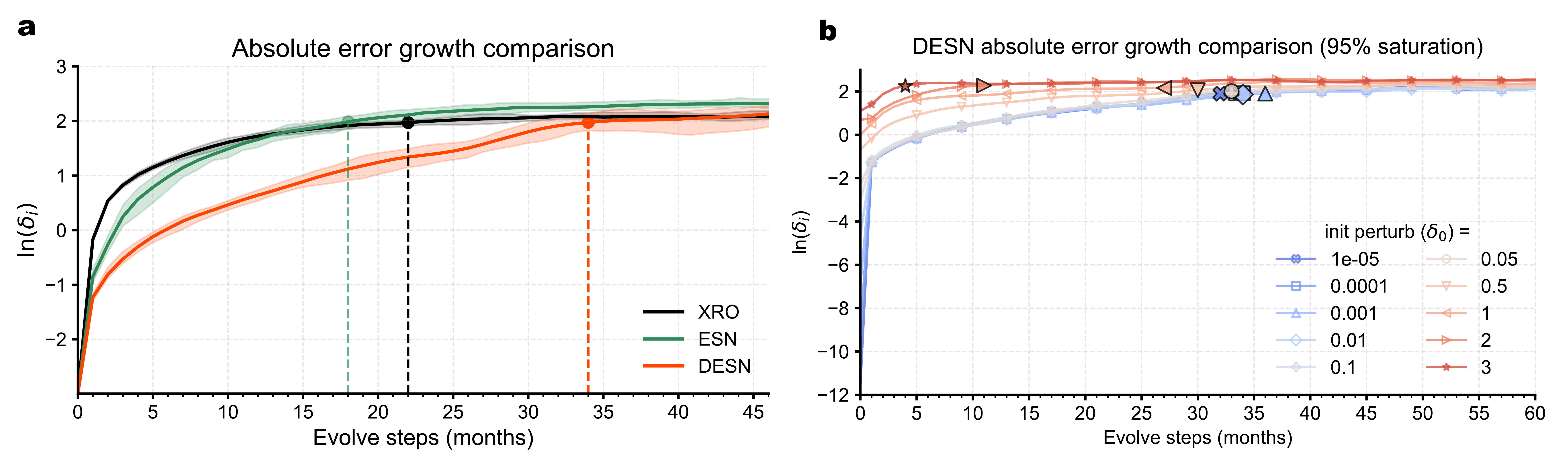}
    \caption{\textbf{Error-growth analysis and nonlinear predictability limits of ENSO models.} 
   \textbf{a},  Evolution of absolute forecast error between perturbed and unperturbed trajectories for the XRO (black), ESN (green), and DESN (red) models, shown as $\ln(\delta_i)$ as a function of lead time. Dashed vertical lines mark the predictability limits for XRO (22 months), ESN (18 months), and DESN (34 months), defined as the time when absolute error reaches 95\% of the XRO saturation value.
   \textbf{b}, Absolute error growth in DESN for a range of initial perturbation magnitudes $\delta_0$. While smaller initial perturbations delay the approach to saturation, the long-term saturation level and the inferred predictability limit remain largely unchanged once $\delta_0 < 0.1$}
    \label{fig:lyapunov}
\end{figure*}

\clearpage

\section*{Methods}\label{sec4}
\subsection{Data and preprocessing}
We use sea surface temperature (SST) and thermocline depth data from the ORAS5 ocean reanalysis dataset (1958–2023) to construct climate indices relevant to ENSO dynamics. Specifically, we derive ten monthly indices from predefined regions (including the Ni\~{n}o 3.4 region and the equatorial Pacific 20 $^\circ$C isotherm depth), covering nine ENSO-related climate modes. These modes were previously identified as physically linked to ENSO variability and are also featured in the extended recharge oscillator (XRO) model~\cite{zhao2024}. The definitions and regions used to calculate these indices are summarized in Supplementary Table. S1.

To isolate ENSO-related variability, we first remove the monthly climatology over the 1979–2010 reference period. We then apply a second-order polynomial detrending to eliminate long-term trends. The resulting standardized anomalies represent interannual variability and are used as model inputs. 

\subsection{Deep Echo State Networks}
Echo State Networks (ESNs) leverage the richness of reservoir dynamics to replicate the temporal patterns expressed in training data. Deep Echo State Networks (DESNs), first proposed by Gallicchio et al.~\cite{gallicchio2017, gallicchio2018}, extend classical ESNs by stacking multiple recurrent reservoir layers. 

\subsubsection{Network Architecture and Dynamics}
At each time step $t$, the DESN processes information hierarchically: the first reservoir layer receives external climate inputs $\mathbf{X}_{\text{in}, t}$, while subsequent layers take as input the state vector from the preceding layer. Let $n_l$ denote the \textit{number of reservoir layers}, and $N_l$ the \textit{number of neurons in layer} $l$---{both are architectural hyperparameters set prior to training (see Table. S2 for specific configurations). We denote the reservoir state of layer $l$ at time $t$ as $\mathbf{r}_t^l \in \mathbb{R}^{N_l}$. 

To emulate the stochastic nature of real climate dynamics, we introduce additive white noise $\boldsymbol{\xi}_{\text{rc}}^l$ with \textit{noise level} $\sigma_{\text{rc}}$ into each reservoir layer. The state update rule for layer $l$ is given by:
\begin{equation}\label{Eq.Dynamcis_DESN} 
 \mathbf{r}_{t+1}^l = (1 - \alpha^l) \mathbf{r}_t^l + \alpha^l \tanh\left(\mathbf{W}_{\text{in}}^l \mathbf{X}_t^l + \mathbf{W}_{\text{res}}^l \mathbf{r}_t^l\right) + \boldsymbol{\xi}_{\text{rc}}^l,
\end{equation}  
where $\alpha^l \in (0,1]$ is the \textit{leak rate} controlling the time-delay property of dynamics in layer $l$, and $\tanh$ is the hyperbolic tangent activation function ensuring bounded neuron activations. The input weight matrix $\mathbf{W}_{\text{in}}^l \in \mathbb{R}^{N_l \times \text{dim}(\mathbf{X}_t^l)}$ is randomly initialized from a uniform distribution and scaled by the \textit{input scaling} parameter $\sigma_{\text{in}}^l$ to control the strength of external inputs. The recurrent weight matrix $\mathbf{W}_{\text{res}}^l \in \mathbb{R}^{N_l \times N_l}$ is constructed as a sparse random matrix with \textit{spectral radius} $\rho^l$ (the largest absolute eigenvalue) and \textit{connection density} $d_{\text{rc}}^l$ (the fraction of nonzero elements) to maintain the echo state property~\cite{jaeger2004}. The layer input $\mathbf{X}_t^l$ is defined hierarchically as:
\begin{equation}\label{Eq.Xin} 
\mathbf{X}^l_t=\left\{\begin{matrix}
\mathbf{X}_{\text{in}, t},\quad & l=1,\\
\mathbf{r}_t^{l-1},\quad & l>1.
\end{matrix}\right.
\end{equation}

\subsubsection{Input Configuration}
In this study, the external input $\mathbf{X}_{\text{in}, t}$ to the first reservoir layer comprises a comprehensive set of climate indices and seasonal cycles:
\begin{equation}
\mathbf{X}_{\text{in}, t}=\begin{bmatrix}
 \text{ENSO}_t & \text{(ENSO index)}\\
 \text{WWV}_t & \text{(Warm Water Volume)}\\
 \text{NPMM}_t & \text{(North Pacific Meridional Mode)}\\
 \text{SPMM}_t & \text{(South Pacific Meridional Mode)}\\
 \text{TNA}_t & \text{(Tropical North Atlantic)}\\
 \text{ATL3}_t & \text{(Atlantic Ni\~{n}o)}\\
 \text{SASD}_t & \text{(South Atlantic Subtropical Dipole)}\\
 \text{IOB}_t & \text{(Indian Ocean Basin)}\\
 \text{IOD}_t & \text{(Indian Ocean Dipole)}\\
 \text{SIOD}_t & \text{(South Indian Ocean Dipole)}\\
\mathbf{PBS}_{t} & \text{(Periodic bootstrap sequence. Eq.~\ref{Eq.PBS})}
\end{bmatrix},
\label{EQ_input}
\end{equation}

\subsubsection{Intrinsic Plasticity for Enhanced Expressiveness}
Following reservoir initialization, we apply the Intrinsic Plasticity (IP) learning rule~\cite{jaegerOptimization2007} to each reservoir unit to enhance network expressiveness. IP optimizes the internal activation dynamics of individual neurons by adapting unit-specific gain ($a$) and bias ($b$) parameters, such that the distribution of activations approximates a target Gaussian distribution with \textit{target mean} $\mu$ and \textit{target standard deviation} $\sigma$. 

This optimization is formulated as minimizing the Kullback–Leibler (KL) divergence between the empirical output distribution and the Gaussian target. For a reservoir unit receiving net input $r_{\text{net}}$, the activation is computed as:
\begin{equation}
\tilde{r} = \tanh(a r_{\text{net}} + b),
\label{EQ_5}
\end{equation}
where the gain $a$ controls the slope of the activation function and the bias $b$ shifts the operating point. At each training step, these parameters are updated according to:
\begin{equation}
\begin{split}
\Delta b &= -\eta \left(-\frac{\mu}{\sigma^2} 
+ \frac{\tilde{r}}{\sigma^2} \left(2\sigma^2 + 1 - \tilde{r}^2 + \mu \tilde{r}\right)\right), \\
\Delta a &= \frac{\eta}{a} + \Delta b \cdot r_{\text{net}},
\end{split}
\label{EQ_6}
\end{equation}
where $\eta$ is the \textit{IP learning rate}. This adaptation process ensures that each reservoir unit maintains a high-entropy, information-rich activation profile, thereby improving the overall dynamical capacity and generalization performance of the DESN.

\subsubsection{Readout Training via Ridge Regression}
After the IP learning process stabilizes the reservoir dynamics, the output of the DESN is computed by a linear readout layer that combines states from all reservoir layers:
\begin{equation}\label{Eq.Wout_DESN}
    \mathbf{\hat{y}}_t = \mathbf{W}_{\text{out}}\begin{bmatrix}
 \mathbf{r}_t^1\\
 \mathbf{r}_t^2\\
 \vdots\\
\mathbf{r}_t^{n_l}
\end{bmatrix},
\end{equation}
where $\mathbf{W}_{\text{out}} \in \mathbb{R}^{N_Y\times\sum_{l=1}^{n_l}N_l}$ is the readout weight matrix connecting all reservoir layers to the output. Unlike the reservoir weights which remain fixed after initialization, $\mathbf{W}_{\text{out}}$ is the only component trained using supervised learning.

We employ ridge regression to determine $\mathbf{W}_{\text{out}}$ by minimizing the regularized least-squares objective:
\begin{equation}
L(\mathbf{W}_{\text{out}}) = \|\mathbf{Y} - \mathbf{W}_{\text{out}} \mathbf{R}\|_2^2 + \lambda \|\mathbf{W}_{\text{out}}\|_2^2,
\end{equation}
where $\mathbf{Y} \in \mathbb{R}^{N_Y \times T}$ is the target matrix containing observed climate states, $\mathbf{R} \in \mathbb{R}^{(\sum_{l=1}^{n_l}N_l) \times T}$ is the concatenated reservoir state matrix across all time steps and layers, and $\lambda \geq 0$ is the \textit{regularization parameter} preventing overfitting. This yields the closed-form solution:
\begin{equation}
\mathbf{W}_{\text{out}} = \mathbf{Y}\mathbf{R}^\top (\mathbf{R}\mathbf{R}^\top + \lambda \mathbf{I})^{-1}.
\end{equation}

In summary, the DESN architecture combines three key components: (1) hierarchical reservoir layers that extract multiscale temporal features with architecture defined by hyperparameters $(n_l, N_l, \alpha^l, \rho^l, d_{\text{rc}}^l, \sigma_{\text{in}}^l, \sigma_{\text{rc}})$ detailed in Table.S2, (2) intrinsic plasticity that optimizes activation distributions with parameters $(\eta, \mu, \sigma)$, and (3) a linear readout trained via ridge regression with regularization $\lambda$. This design maintains computational efficiency while achieving strong generalization for long-lead climate prediction.
\subsection{Training and validation periods}
We employed three complementary experiments to assess model performance,

1. \textit{Out-of-sample forecast}: The model was trained on data from 1958–1999 and validated on 2002–2023, ensuring that the validation set is temporally independent of the training period and thus testing the model’s ability to generalize to new periods.

2. \textit{In-sample forecast}: The model was both trained and validated on the 1979–2023 dataset. This experiment evaluates the model’s ability to reproduce known data, providing insight into its fitting capacity.

3. \textit{Cross-validation}: The full 1958–2023 dataset was divided into 11 subsets of 6 consecutive years. For each iteration, 24 consecutive years were withheld for validation, with the remainder used for training. This rolling-block design ensures that all portions of the record contribute to both training and validation, yielding a robust evaluation less dependent on specific temporal divisions.

The Table.~\ref{tab:validation_methods} summarizes these validation settings.
\begin{table}[h!]
\centering
\renewcommand{\arraystretch}{1.4}
\caption{Validation settings for model performance assessment.}
\label{tab:validation_methods}
\begin{tabular}{lll} \hline 
\textbf{Validation method} & \textbf{Training period} & \textbf{Validation period} \\ \hline 
Out-of-sample forecast & 1958--1999 & 2002--2023 \\  
In-sample forecast & 1979--2023 & 1979--2023 \\  
Cross-validation & Remaining 7 subsets \newline (e.g., 1958--1975, 2000--2023) & 4 consecutive subsets \newline (e.g., 1976--1999) \\ 
\hline 
\end{tabular}
\end{table}

\subsection{Prediction skill metrics}
Forecast skill was evaluated using two standard metrics: the ACC and the root mean square error (RMSE). The ACC, calculated as the Pearson correlation between forecasts ($f$) and observations ($o$),
\begin{equation}
\text{ACC} = \frac{\sum (f - \bar{f})(o - \bar{o})}{\sqrt{\sum (f - \bar{f})^2 \sum (o - \bar{o})^2}},
\end{equation}
where $\bar{f}$ and $\bar{o}$ are the means of the forecasts and observations, respectively.  The RMSE measures the average forecast error magnitude,
\begin{equation}
\text{RMSE} = \sqrt{\overline{(f - o)^2}},
\end{equation}
where the overbar denotes the mean.

\subsection{Quantitative Experiments for Predictability Source}\label{subsec:predictability_sources}

To rigorously quantify the contribution of individual climate modes and their interactions to ENSO predictability, we conducted four complementary sets of systematic experiments. All experiments use the same training (1958--1999) and validation (2002--2024) periods.

\subsubsection{Mode Decoupling Experiments}
Beyond Niño~3.4, we consider nine supplementary climate indices relevant to ENSO dynamics~\cite{zhao2024} (Fig.~\ref{fig1}a), forming a ten-variable predictor set in the full model. Mode-decoupling experiments are performed by removing one supplementary mode $j$ at a time from the full predictor set, while keeping all remaining inputs, model architecture, and training procedure unchanged.

The full model, trained using all ten climate indices (Eq.~\eqref{EQ_input}), serves as the control experiment. For example, DESN-$D_{\mathrm{NPMM}}$ denotes a DESN experiment in which NPMM is excluded from both training and prediction, with all other inputs and model configurations identical to the full model. Analogous experiments are performed for the XRO framework, denoted as XRO-$D_j$ (e.g., XRO-$D_{\mathrm{WWV}}$, XRO-$D_{\mathrm{NPMM}}$). The difference in forecast skill between the full model and $D_j$ quantifies the sensitivity of ENSO predictability to the removal of climate mode $j$. Comparing DESN-$D_j$ with XRO-$D_j$ under identical predictor configurations reveals the role of nonlinear interactions in cross-basin interactions.

\subsubsection{Mode Addition Experiments}

The baseline configuration consists of Niño~3.4 and WWV, representing the core ENSO recharge–discharge subsystem. Mode-addition experiments are performed by adding one supplementary climate mode $j$ to this baseline, while keeping all other model settings unchanged.

For example, DESN-$A_{\mathrm{NPMM}}$ denotes a DESN configuration in which Niño~3.4, WWV, and the North Pacific Meridional Mode (NPMM) are used as inputs. Identical experiments are conducted for the XRO framework, denoted as XRO-$A_j$ (e.g., XRO-$A_{\mathrm{NPMM}}$, XRO-$A_{\mathrm{SPMM}}$).

The difference in forecast skill between $A_j$ and the baseline configuration quantifies the sensitivity of ENSO predictability to the inclusion of climate mode $j$ beyond the core recharge–discharge dynamics. Because XRO is governed primarily by linear dynamics with prescribed coupling structure, systematic skill differences between DESN-$A_j$ and XRO-$A_j$ under identical inputs highlight processes that require nonlinear representations beyond linear superposition.

\subsubsection{Increasing input-dimension experiments}
The baseline configuration ($d=2$) consists of Niño~3.4 and WWV, representing the minimal recharge–discharge system. Higher-dimensional configurations are constructed by sequentially adding supplementary climate modes to this baseline, yielding input dimensions $d = 3, 4, \ldots, 10$. For each dimension $d$, all possible combinations of $d-2$ modes are selected from the remaining eight supplementary climate indices, while Niño~3.4 and WWV are always retained. Consequently, the $d=10$ configuration corresponds to the fully coupled model.

For each input dimension, we compute the mean forecast skill by averaging over all corresponding sub-model realizations. This approach isolates the systematic effect of increasing coupling dimensionality, independent of the influence of any single climate mode. Identical increasing-dimension experiments are conducted for both DESN and XRO frameworks.

\subsubsection{Uninitialized experiments}

To investigate state-dependent predictability and the role of initial-condition memory, we performed a set of uninitialized experiments, denoted as $U_j$. In these experiments, the initial condition of climate mode $j$ is replaced by its climatological mean (zero in standardized units), while all other initial conditions are retained as observed.

Importantly, all uninitialized experiments are conducted using the fully trained DESN model, which is trained with the complete set of coupled climate modes. Thus, the learned interaction structure remains unchanged, and only the initial-state information of mode $j$ is removed. The impact of the initial condition of mode $j$ is quantified by the difference in forecast skill between the control experiment and $U_j$.

To further disentangle the role of basin-scale memory, we also conducted ocean-level uninitialized experiments by simultaneously replacing the initial conditions of all climate modes within a given ocean basin by climatology. Specifically, $U_{\mathrm{NPMM+SPMM}}$ removes the initial-state information of both Pacific Meridional Modes, $U_{\mathrm{IOB+IOD+SIOD}}$ removes all Indian Ocean modes, and $U_{\mathrm{TNA+ATL3+SASD}}$ removes all Atlantic modes.

The resulting differences in forecast skill isolate the contribution of initial-condition information associated with individual modes or entire ocean basins to ENSO predictability. Unlike mode-decoupling experiments, which remove predictors entirely and alter the dynamical system learned during training, uninitialized experiments retain all learned nonlinear interactions and instead diagnose sensitivity to initial-state uncertainty.

\subsection{Sparse-Nonlinear XRO (SN-XRO) model}\label{sec:snxro}

For reference, the state vector of XRO is given by
\begin{equation}
\mathbf{X} =
\begin{pmatrix}
T_{\mathrm{ENSO}},\;
h_{\mathrm{WWV}},\;
T_{\mathrm{NPMM}},\;
T_{\mathrm{SPMM}},\;
T_{\mathrm{IOB}},\;
T_{\mathrm{IOD}},\;
T_{\mathrm{SIOD}},\;
T_{\mathrm{TNA}},\;
T_{\mathrm{ATL3}},\;
T_{\mathrm{SASD}}
\end{pmatrix}^{\mathrm{T}} .
\end{equation}

In the original XRO formulation, the system evolves according to
\begin{equation}
\frac{d\mathbf{X}}{dt}
=
\mathbf{L}\mathbf{X}
+
\mathbf{N}_{\mathrm{XRO}}(\mathbf{X}),
\end{equation}
where the nonlinear terms are restricted to
\begin{equation}
\mathbf{N}_{\mathrm{XRO}}(\mathbf{X}) =
\begin{pmatrix}
b_1 T_{\mathrm{ENSO}}^2 + b_2 T_{\mathrm{ENSO}} h_{\mathrm{WWV}} \\
0 \\
0 \\
0 \\
b_3 T_{\mathrm{IOD}}^2 \\
0 \\
0 \\
0 \\
0 \\
0
\end{pmatrix}.
\end{equation}
Thus, nonlinear interactions in XRO are confined to ENSO asymmetry, ENSO–WWV recharge–discharge coupling, and a conditional quadratic contribution associated with the Indian Ocean Dipole.

In the SN-XRO model, we extend the nonlinear tendency by introducing additional quadratic coupling terms that involve ENSO and WWV:
\begin{equation}
\mathbf{N}_{\mathrm{SN}}(\mathbf{X}) =\mathbf{N}_{\mathrm{XRO}}(\mathbf{X}) +
\begin{pmatrix}
\textbf{N}_1T_{\mathrm{ENSO}}\textbf{T}_{\mathrm{M}}\\
\textbf{N}_2h_{\mathrm{wwv}}\textbf{T}_{\mathrm{M}}\\
0 \\
0 \\
0 \\
0 \\
0 \\
0 \\
0 \\
0
\end{pmatrix},
\end{equation}
where 
\begin{equation}
\mathbf{T}_{\mathrm{M}} =
\begin{pmatrix}
T_{\mathrm{NPMM}},\;
T_{\mathrm{SPMM}},\;
T_{\mathrm{IOB}},\;
T_{\mathrm{IOD}},\;
T_{\mathrm{SIOD}},\;
T_{\mathrm{TNA}},\;
T_{\mathrm{ATL3}},\;
T_{\mathrm{SASD}}
\end{pmatrix}^{\mathrm{T}},
\end{equation}
denotes the set of external climate modes. No nonlinear interaction terms are introduced among the external climate modes themselves. 

Here, $\mathbf{N}_1$ and $\mathbf{N}_2$ are seasonally modulated coefficient vectors controlling the nonlinear coupling between ENSO, WWV, and inter-basin climate modes. Each coefficient is parameterized as a truncated Fourier series,
\begin{equation}
n(t)
=
n_0
+
n_{1s}\sin(\omega t)
+
n_{1c}\cos(\omega t)
+
n_{2s}\sin(2\omega t)
+
n_{2c}\cos(2\omega t),
\end{equation}
where $\omega = 2\pi/12$ represents the annual cycle.

The Sparse Identification of Nonlinear Dynamical systems (SINDy) method~\cite{brunton2016,desilva2020} is applied to fit the parameters of the prescribed quadratic interaction terms using ORAS5 data over the training period (1979--2001). This procedure automatically eliminates statistically insignificant nonlinear terms, yielding a parsimonious representation of nonlinear coupling consistent with the data. The resulting seasonally modulated nonlinear coupling parameters are shown in Fig.~S9.

To ensure a controlled comparison between XRO and SN-XRO, model parameters are estimated in two successive steps.

First, the standard XRO model is fitted using the SINDy framework, including all linear coefficients $L_{ij}$ and the prescribed nonlinear terms $b_1 T_{\mathrm{ENSO}}^2$, $b_2 T_{\mathrm{ENSO}} h_{\mathrm{WWV}}$, and $b_3 T_{\mathrm{IOD}}^2$. This step yields a physically consistent baseline representation of ENSO and its linear cross-basin couplings.

In the SN-XRO model, all linear coefficients $L_{ij}$ are fixed to the values obtained from the XRO fitting and remain unchanged. Additional nonlinear terms are then  estimated using SINDy.

This hierarchical fitting strategy ensures that differences between XRO and SN-XRO arise solely from enhanced nonlinear coupling involving ENSO and WWV, rather than from changes in linear dynamics.

SN-XRO is integrated deterministically and evaluated over the same out-of-sample period (2002--2024) as XRO and DESN, ensuring that differences in forecast performance arise solely from the inclusion of physically constrained nonlinear interaction pathways.

By construction, SN-XRO provides a minimal nonlinear extension of XRO that allows us to directly test whether enhanced ENSO–WWV–inter-basin coupling is sufficient to reproduce the delayed error growth and extended predictability identified in DESN.

\subsection{Computation of Nonlinear Local Lyapunov Exponents}

The XRO, ESN, and DESN are all autonomous dynamical systems whose forward evolution can be expressed as $\mathbf{z}_{t+1} = \mathbf{F}(\mathbf{z}_t)$, where $\mathbf{z}_t$ is the full state vector and $\mathbf{F}$ is the evolution operator. The state space differs across models:
\begin{equation}
\mathbf{z}_t = \begin{cases}
\mathbf{X}_t \in \mathbb{R}^{10}, & \text{XRO} \\
[\mathbf{r}_t; \mathbf{X}_t] \in \mathbb{R}^{N+10}, & \text{ESN} \; (N = 20000) \\
[\mathbf{r}^1_t; \mathbf{r}^2_t; \mathbf{X}_t] \in \mathbb{R}^{N_1+N_2+10}, & \text{DESN} \; (N_1 = 20000, N_2 = 12000)
\end{cases}
\end{equation}
where $\mathbf{X}_t \in \mathbb{R}^{10}$ denotes the 10 climate indices, and $\mathbf{r}_t$, $\mathbf{r}^1_t$, $\mathbf{r}^2_t$ are the reservoir neuron states.

Starting from an initial climate state $\mathbf{X}_0$ on the attractor, we introduce a small perturbation to the 10 climate indices: $\mathbf{X}'_0 = \mathbf{X}_0 + \boldsymbol{\delta}_0$, where $\boldsymbol{\delta}_0$ is a random unit vector scaled by initial amplitude $\epsilon_0$. For ESN/DESN, the corresponding reservoir states are initialized from the unperturbed climate state, ensuring perturbations only affect the climate variables. Both the reference trajectory $\mathbf{X}_i$ and perturbed trajectory $\mathbf{X}'_i$ are evolved independently using the full nonlinear dynamics for $T = 100$ months.

We test a range of initial perturbation amplitudes $\epsilon_0$. For each amplitude, we perform 500 realizations with different random perturbation directions, sampled from 500 initial conditions uniformly distributed across the test period (2002--2023).

The error at step $i$ is defined as:
\begin{equation}
\delta_i = \|\mathbf{X}'_i - \mathbf{X}_i\|_2.
\end{equation}
where $\|\cdot\|_2$ denotes the Euclidean norm (L2 norm). The absolute error growth is quantified by the logarithmic error:
\begin{equation}
E_i = \ln \delta_i.
\end{equation}

To define a consistent saturation threshold across models with different error growth rates, we compute the mean logarithmic error over months 60--100 of XRO evolution:
\begin{equation}
E_{\text{sat}} = \frac{1}{40} \sum_{i=60}^{100} E_{\text{XRO},i}.
\end{equation}

The predictability horizon $T_p$ is defined as the first time when the 95 percentile of $E_{\text{sat}}$ is reached:
\begin{equation}
T_p = \min \left\{ i \mid \langle E_i \rangle \geq \gamma E_{\text{sat}} \right\},
\end{equation}
where $\langle \cdot \rangle$ denotes averge over all reanlization, and $\gamma =0.95$ is the saturation threshold.

\section*{Data availability}
Datasets and models used in this paper are freely available. ORAS5: \url{https://cds.climate.copernicus.eu/datasets/reanalysis-oras5?tab=overview}; XRO model ENSO forecast: \url{https://github.com/senclimate/XRO}; 3D-Geoformer
model ENSO forecast: \url{https://msdc.qdio.ac.cn/data/metadata-special-detail?id=1602252663859298305&otherId=1602252664035459074}; NMME: \url{https://iridl.ldeo.columbia.edu/SOURCES/.Models/.NMME/}; IRI ENSO forecast: \url{https://iri.columbia.edu/our-expertise/climate/forecasts/enso/current/} ;
Ni\~{n}o3.4 Index from the HadISST1.1: \url{https://psl.noaa.gov/data/timeseries/month/DS/Nino34/}.

\section*{Code availability}
The Python codes used for the analysis are available on GitHub (\url{https://github.com/zhangzejing/RC-ENSO}).

\section*{Funding Declaration}
This work was supported by the National Natural Science Foundation of China (Grant No. 42575057, T2525011, 42450183, 12275020, 12135003, 12205025, 42461144209, 62333002), the Ministry of Science and Technology of China (2023YFE0109000), 
the National Key R\&D Program of China (2025YFF0517203, 2025YFF0517304),
and  State Key Laboratory of Information Photonics and Optical Communications (IPOC2023ZJ02). J.F. is supported by the Fundamental Research Funds for the Central Universities.

{\section*{Author Contributions}
J.M, J.X and J.F designed the research. Z.Z performed the analysis. Z.Z, J.M, Z.Q, W.D, J.G, Z.Y, J.X, X.C, W.C, J.K, S.H and J.F prepared the manuscript, discussed results, and contributed to writing the manuscript. J.M and J.F led the writing of the manuscript. }

\section*{Additional information}
Supplementary Information is available in the online version of the paper.

\section*{Competing interests}
The authors declare no competing interests.

\end{document}